\documentclass[reprint,aps,prb,superscriptaddress,amsmath]{revtex4-2}
\usepackage{amsmath}
\usepackage{amsfonts}
\usepackage[normalem]{ulem} % AW
\usepackage{braket} 
\usepackage{xcolor}
\usepackage{appendix}
\usepackage{enumitem}
\usepackage{graphicx}
\usepackage{float}
\RequirePackage[hyperindex,colorlinks,bookmarksnumbered, plainpages=true,pdfstartview=FitH]{hyperref}
\hypersetup{linkcolor=blue,urlcolor=blue,citecolor=blue}
\usepackage{hyperref}

	\newcommand{\Sent}{S_{\mathrm{ent}}}
	\definecolor{darkgreen}{rgb}{0.01, 0.75, 0.24}
	\usepackage{soul}
\newcommand{\Eq}[1]{Eq.~\eqref{#1}}

\newcommand{\Fig}[1]{Fig.~\ref{#1}}
\newcommand{\Figs}[1]{Figs.~\ref{#1}}

\begin{document}
\sloppy

\title{Finite-size subthermal regime in disordered SU$(N)$-symmetric Heisenberg chains}

\author{Dimitris Saraidaris}
\affiliation{Arnold Sommerfeld Center for Theoretical Physics, Center for NanoScience,\looseness=-1\,  and Munich 
Center for \\ Quantum Science and Technology,\looseness=-2\, Ludwig-Maximilians-Universität München, 80333 Munich, Germany}
\affiliation{Department of Physics,\looseness=-2\, Freie Universität Berlin, 14195 Berlin, Germany}
\author{Jheng-Wei Li}
\affiliation{Arnold Sommerfeld Center for Theoretical Physics, Center for NanoScience,\looseness=-1\,  and Munich 
Center for \\ Quantum Science and Technology,\looseness=-2\, Ludwig-Maximilians-Universität München, 80333 Munich, Germany}
\affiliation{Universit{\'e} Grenoble Alpes, CEA, Grenoble INP, IRIG, Pheliqs, F-38000 Grenoble, France}
\author{Andreas Weichselbaum}
\affiliation{Department of Condensed Matter Physics and Materials Science, Brookhaven National Laboratory, Upton, New York 11973-5000, USA}
\author{Jan von Delft}
\affiliation{Arnold Sommerfeld Center for Theoretical Physics, Center for NanoScience,\looseness=-1\,  and Munich 
Center for \\ Quantum Science and Technology,\looseness=-2\, Ludwig-Maximilians-Universität München, 80333 Munich, Germany}
\author{Dmitry A. Abanin}
\affiliation{Department of Theoretical Physics, University of Geneva, 1211 Geneva, Switzerland}
\affiliation{Department of Physics, Princeton university, Princeton, New Jersey 08544, USA}

\begin{abstract}
SU$(N)$ symmetry is incompatible with the many-body localized (MBL) phase, even when strong disorder is present. However, recent studies have shown that finite-size SU$(2)$ systems exhibit non-ergodic, subthermal behavior, characterized by the breakdown of the eigenstate thermalization hypothesis, and by the excited eigenstates entanglement entropy that is intermediate between area and volume law. 
In this paper, we extend previous studies of the SU$(2)$-symmetric disordered Heisenberg model to larger systems, using the time-dependent density matrix renormalization group (tDMRG) method. We simulate quench dynamics from weakly entangled initial states up to long times, finding robust subthermal behavior at stronger disorder. Although we find an increased tendency towards thermalization at larger system sizes, the subthermal regime persists at intermediate time scales, nevertheless, and therefore should be accessible experimentally.
At weaker disorder, we observe signatures of thermalization, however, entanglement entropy exhibits slow sublinear growth, in contrast to conventional thermalizing systems. 
Furthermore, we study dynamics of the SU$(3)$-symmetric disordered Heisenberg model. Similarly, strong disorder drives the system into subthermal regime, albeit thermalizing phase is broader compared to the SU$(2)$ case. Our findings demonstrate the robustness of the subthermal regime in spin chains with non-Abelian continuous symmetry, and are consistent with eventual thermalization at large system sizes and long time scales, suggested by previous studies. 

\end{abstract}
\date{\today}

\maketitle

\newpage

\section{Introduction} 

The non-equilibrium dynamics of interacting isolated systems has recently drawn a lot of attention, both due to its theoretical significance and experimental applications. In the absence of an external bath, system's dynamics is determined by the intrinsic interactions between its constituents. Ergodic isolated systems, for which all microstates are accessible, evolve towards thermal equilibrium. Such a system itself acts as a thermal bath for its subsystems, as long as they are small enough. In this case, the local observables of the system reach their thermal expectation values at long enough times for arbitrary physical initial states -- the behavior that is understood in terms of the celebrated Eigenstate Thermalization Hypothesis (ETH)~\cite{Deutsch1991,Srednicki1994,Rigol2008}.

The presence of strong disorder can drastically change this behavior, and lead to a breakdown of ergodicity. The first evidence of localization due to disorder is described in the seminal paper of P.W. Anderson \cite{Anderson1958}, the so-called Anderson localization, referring to a single-particle localization. Furthermore, recent theoretical \cite{Basko2006, Gornyi2005, Oganesyan2007, Znidaric2008, Vosk2013, Serbyn2013, Serbyn2013_2, Huse2014, Bardarson2012, Ros2015, Ponte2015, Abanin2016, Lazarides2015} and experimental studies \cite{Bloch2008, Blatt2012, Schreiber2015, Ovadia2015, Smith2016, Choi2016, Choi2017, Bordia2017, Xu2018, Lukin2019, Rispoli2019} of isolated systems have discovered that strong disorder can suppress thermalization in a many-body setting. Thus, a recently emerged phenomenon, the many-body localization (MBL) (for reviews, see \cite{Alet2018, Abanin2019}) and its transitions \cite{Pal2010, Torres2015, Vosk2015, Potter2015, Devakul2015, Lim2016, Singh2016, Panda_2019}, have drawn much attention. An MBL phase constitutes a new dynamical phase of matter, where ergodicity is broken and information about the initial state is preserved throughout the evolution of the system. The key characteristic of the MBL phase is that such systems exhibit a complete set of quasi-local integrals of motion (LIOMs) \cite{Serbyn2013_2, Huse2014}. This leads to an area law scaling of the entanglement entropy
$S(\ell)=-\mathrm{Tr}\left(\rho^{(\ell)}\ln\rho^{(\ell)}\right)$ for a contiguous block of size $\ell$ for
excited eigenstates, as well as a logarithmic growth of entanglement entropy with time in a quantum quench \cite{Znidaric2008, Vosk2013, Serbyn2013, Serbyn2013_2, Huse2014, Bardarson2012}. 

While the MBL phenomenology in finite-size systems has been firmly established, in a recent exact diagonalization study \cite{Suntajs2020}, the stability of the MBL phase in 1d in the limit where both system size and time go to infinity was challenged, based on a specific extrapolation of the finite-size numerical results to this limit. This interpretation was however questioned in subsequent papers, see  Refs. ~\cite{Abanin2021,Sierant20}. Furthermore, Ref.~\cite{Panda_2019} argued that accessible system sizes may not be sufficient to draw conclusions regarding the  MBL-thermal transition (or its absence). We note that in 2d, there are arguments suggesting that rare thermal inclusions lead to extremely slow thermalization of the system~\cite{HuveneersPRB2017}. The avalanche instability is  however not effective for 1d systems at strong disorder. We emphasize that the important issues of eventual stability of the MBL phase are beyond the scope of this paper.

In this paper, we address the interplay between SU$(N)$ symmetry, thermalization, and localization. To that end, we investigate whether disordered, SU$(N)$ symmetric Heisenberg chains thermalize or instead show signatures of ergodicity breaking. The motivation for our study is twofold. Firstly, previous studies established that ergodicity breaking depends on the symmetry of a disordered system. While Abelian symmetries allow the existence of the MBL phase, with the example of the Heisenberg chain in a random magnetic field being arguably the most well-studied system in the MBL phase \cite{Znidaric2008, Pal2010, Luitz2015, Serbyn2016}, {\it non-Abelian symmetries} impose strong constraints on ergodicity breaking~\cite{Potter2016}. For models with discrete non-Abelian symmetries, two scenarios have been discussed. The first possibility is that such systems form an MBL phase, in which the symmetry is spontaneously broken down to an Abelian subgroup \cite{Vasseur2016}. Alternatively, they may exhibit a symmetry-preserving quantum critical glass (QCG) \cite{Vasseur2015}, however this latter phase was recently argued to be perturbatively unstable to the proliferation of resonances, which lead to an eventual thermalization~\cite{WarePRB2021}. Continuous non-Abelian symmetry groups with infinite-dimensional representations were argued to be inconsistent with the standard MBL with a complete set of LIOMs~\cite{Potter2016, Thomson2022}. 

Secondly, in a recent study \cite{Abanin2020} of disordered SU$(2)$-symmetric Heisenberg systems it was found that when disorder is sufficiently strong, an intermediate regime emerges, which is neither fully MBL nor thermal. This regime, which we refer to as \textit{subthermal} below, exhibits entanglement scaling of excited eigenstates that is intermediate between the area and volume law. In this paper, the authors performed an exact diagonalization (ED) analysis of chains up to $L\leq 24$ sites, and observed that the behaviour of the systems is clearly subthermal. Furthermore, a strong-disorder renormalization group (SDRG) analysis revealed that long-range resonances eventually proliferate, leading to the breakdown of non-ergodic structure of eigenstates for very large systems, even subject to very strong disorder.

In this paper, we extend the work in Ref.~\cite{Abanin2020} by studying the dynamical properties of disordered SU$(N)$-symmetric Heisenberg model. Specifically, we check if subthermal behavior can appear for larger systems and also in SU$(3)$ symmetric models. The analysis is based on the time evolution of short-range entangled initial states, expressed in the form of matrix product states (MPS), under the disordered Heisenberg Hamiltonian. The evolution method we use is the time-dependent density matrix renormalization group (tDMRG) \cite{Daley2004, White2004, Weichselbaum2012}. We compute the equal-time spin-spin correlations and entanglement entropy of the resulting time-evolved states, and analyze their scaling with length and evolution in time.

\section{Model and Methods} 

We consider the disordered SU$(N)$-symmetric Heisenberg model \footnote{For a study of disordered SU$(N)$ chains using exact diagonalization, see B. Dabholkar and F. Alet, in preparation.} for $N=2$ and $3$ particle flavors, described by the Hamiltonian,
\begin{align}
 H = \sum_{i=1}^{L-1} % L % AW: L here would suggest periodic BC
 J_i\,\mathbf{S}_i \cdot \mathbf{S}_{i+1}.
 \label{eq:HAM}
\end{align}
Here, $\mathbf{S}_i$ are the SU$(N)$ generators in their fundamental representations on site $i$ of a one-dimensional chain of length $L$ with open boundary conditions.
The nearest-neighbor couplings are chosen antiferromagnetic, throughout, with the values randomly drawn from the normalized power-law probability distribution
\begin{align}
P(J)=\frac{\alpha }{J^{1-\alpha}} \, \theta (1-J) \qquad (0<J<1)
\label{eq:distribution}
\end{align}
in accordance with early theoretical studies \cite{Theodorou1976, Ma1979, Dasgupta1980} and experiments \cite{Scott1975}. 
The real parameter $\alpha > 0$ controls the strength of the disorder, where plain random couplings $J\in[0,1]$
are obtained for $\alpha=1$, and uniform couplings
for $\alpha\to\infty$.
The unit of energy is set by choosing the upper cutoff $J<1$.
The average ratio of the larger over the smaller value of a pair of sampled coupling constants $J_1 < J_2$ diverges for $\alpha\to 0$, $\langle \frac{J_2}{J_1}\rangle \to\infty$ (and it has a typical value of  $\exp \langle \ln \frac{J_2}{J_1}\rangle
= e^{1/\alpha}$\cite{Abanin2020} which, while smooth across $\alpha=1$,
also % it
diverges rapidly for $\alpha < 1$). This demonstrates the strong non-uniformity of couplings along the 
chain for the strong disorder regime $\alpha<1$, which also leads to weakly coupled bonds along the chain. 

In this paper, we focus on the \textit{strong} disorder region $\alpha \leq 1$, where for $N=2$ subthermal behavior for finite spin chains has been reported \cite{Abanin2020}. To verify the advocated subthermal behavior, we use the tDMRG to study the real-time dynamics of weakly entangled initial states whose energy is close to zero. Such states lie in the middle of the many-body spectrum. The advantage of our approach is twofold. First, for disordered systems, tDMRG is a well-controlled exact MPS approach that allows us to simulate system sizes much larger than those accessible to exact diagonalization \cite{Chanda2020}
(we study system sizes up to $L=144$). Second, within our MPS framework, we can not only study the relaxation of local observables in real-time, but also measure the scaling of the averaged entanglement entropies of small subsystems straightforwardly \cite{Yu2016}.
Both are important indicators of tendencies towards thermalization, sub-thermalization, or MBL.

Below, we first perform real-time simulations of given initial states for several disorder realizations via tDMRG with the second-order Trotter decomposition, time step $dt=0.1$ 
and {the effective bond dimension of $D^\ast=2048$ multiplets}.
We use the resulting time-evolved MPSs to study the scaling of the averaged entanglement entropies for small subsystems and the nearest neighbor correlators $\braket{\mathbf{S}_{i}\cdot \mathbf{S}_{i+1}}$ in real-time as well as in the long time limit. As discussed below, the decay of the correlator to a thermal value provides a signature of thermalization, while its non-thermal value signals a non-thermal regime.
In order to extract systematic behavior, we will vary both
the system size $L$ and the disorder strength $\alpha$ for the SU$(2)$
as well as the SU$(3)$ symmetric model.

\section{SU$(2)$-symmetric model} 

\begin{figure}[bt!]
\includegraphics[width=1\linewidth,trim = 0.0in 0.0in 0.0in 0.0in,clip=true]{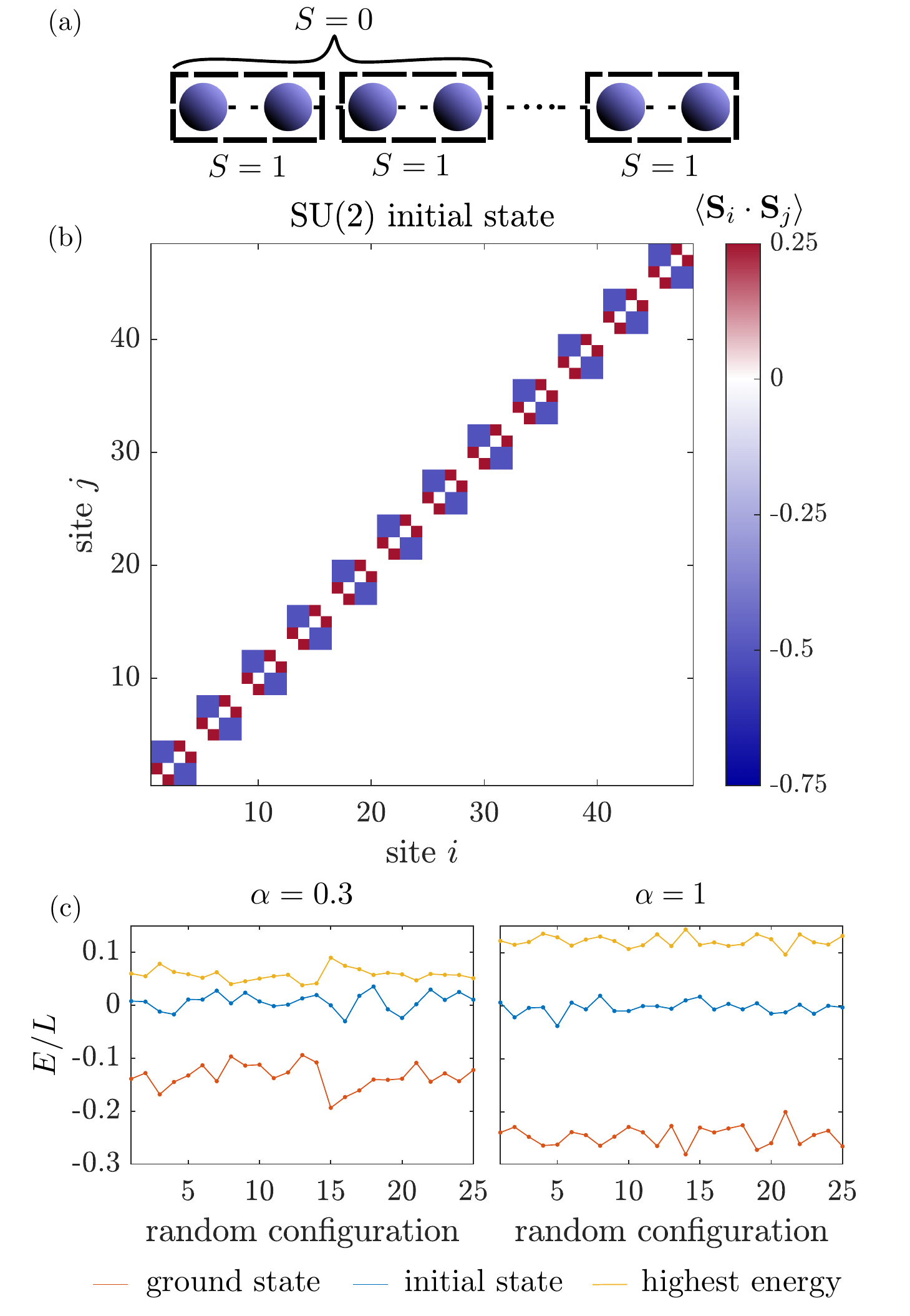}
  \caption{(a) A schematic depiction of the structure of the initial state for the SU$(2)$ model: neighboring spins form triplets on odd bonds, and pairs of neighboring triplets are coupled to form singlets. (b) Spin-spin correlations $\langle\textbf{S}_i\cdot\textbf{S}_j\rangle$ color-plots for the SU$(2)$ initial state. Correlations between spins forming triplets or singlets yield $\langle\textbf{S}_i\cdot\textbf{S}_j\rangle = +0.25$ (dark red) or $-0.75$ (dark blue) respectively.
  (c) Ground state (GS), initial state, and highest energy of $25$ random configurations for disorder strength $\alpha=0.3$ (left) and $1$ (right),
  where $E_{\mathrm{highest}}(J)=-E_{\mathrm{GS}}(-J)$.
}
  \label{fig:initial_states}
\end{figure}

In this section, we focus on the disordered SU$(2)$ Heisenberg model. First, we introduce an initial state and study its time evolution. Second, we present the results of the entanglement entropy and spin-spin correlations measurements.

\subsection{Initial state}

In the context of thermalization, the choice of initial states for the real-time evolution is important because different parts of the energy spectrum can exhibit qualitatively different dynamics.
When disorder is introduced, the states near both edges of the energy spectrum generally localize before the states in the middle of the spectrum, where the density of states is high.
If such transitions do occur as a function of energy density, it signifies the appearance of mobility edges
that separate thermal and MBL states in the energy spectrum.
In this paper, we shall restrict our attention to the middle of the spectrum by preparing initial states with energies close to zero. Such states are expected to exhibit the strongest tendency to thermalization. Furthermore, since the total spin is a conserved quantity, hereafter we focus on the singlet sector with $S_{\rm{tot}} = 0$. Such a choice is relevant for the experimental setup of strongly interacting fermions at half-filling \cite{Schreiber2015}.

Concretely, we always prepare the initial MPS $\ket{\psi_{t=0}}$ as depicted in \Fig{fig:initial_states}(a,b): 
Every pair of nearest-neighbor spins on odd bonds is combined into a triplet, $\textbf{S}=1$, with two neighboring such triplets then fused
into a spin singlet. Thus the total spin of a chain with length of a multiple of $4$ is zero. The initial MPS prepared in this way has a small bond dimension. Importantly, the energy of this initial state is always close to zero for any disorder realizations, as shown in \Fig{fig:initial_states}(c).

\subsection{Entanglement entropy scaling}

\begin{figure*}[!http]
\includegraphics[width=1\textwidth]{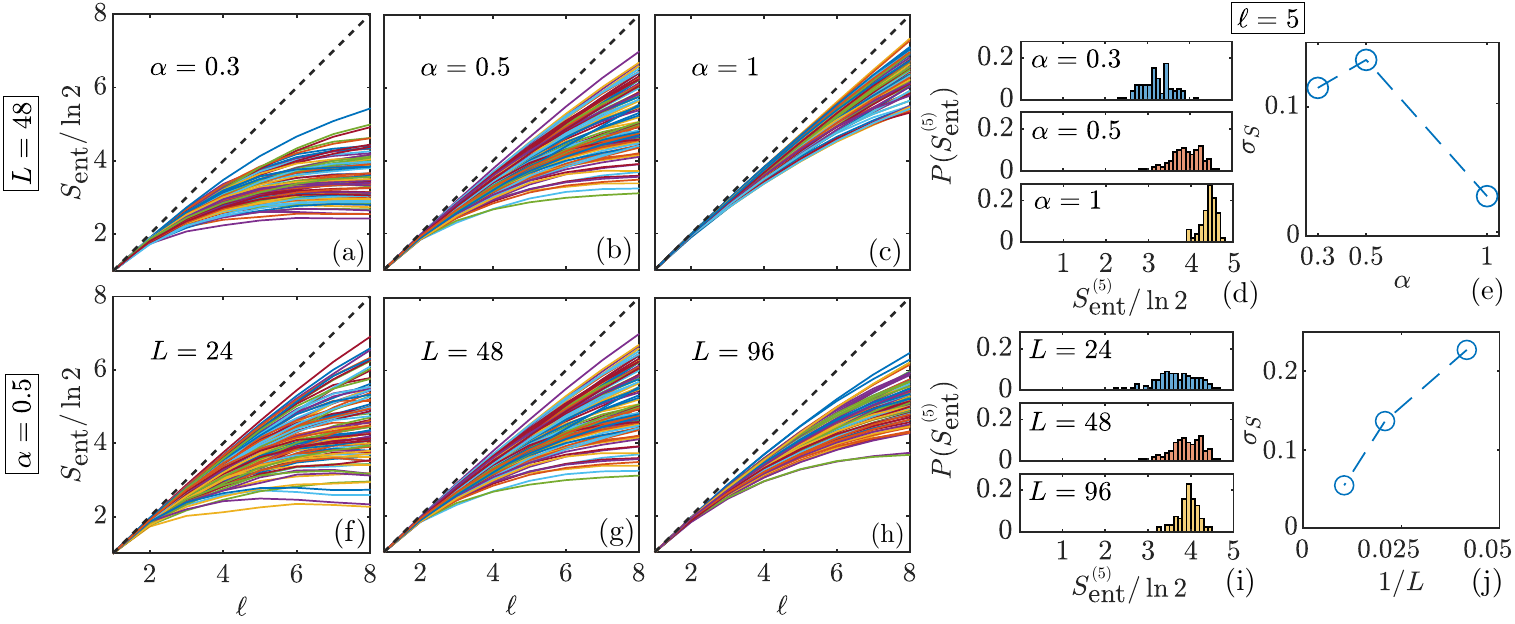}
\caption{
   Entanglement entropy $\Sent^{(\ell)}$ at final evolution time
   $t_\mathrm{f}$ for various combinations of the system size $L$,
   disorder strength $\alpha$ and subsystem size $\ell$. We
   chose $t_\mathrm{f}=500$ for all cases, % considered
   except for
   $\alpha = 1$, $L=48$, for which we used $t_\mathrm{f} =150$
   due to the high computational cost of longer times.
   Top row: fixed $L=48$, $\alpha\in \{0.3, 0.5, 1  \}$.
   Bottom row:
   fixed $\alpha = 0.5$, $L\in \{24, 48, 96\}$ (for ease of comparison, hence the central panels (b) and (g) are identical). (a-c, f-h):
   Dependence of $\Sent$ on $\ell$ where each line represents a
   different disorder configuration,
    showing 100 disorder configurations in each panel.
   (d,i): Distribution of
   $\Sent$ values for $\ell=5$.
   (e,j): Variance of the
   distributions from (d,i).
   }
  \label{fig:caee}
\end{figure*}

The behavior of the entanglement entropy
at a subsystem level is a usefull indicator of the thermal or localized fate of the system.
In particular, if the system 
is (close to) thermalized at the final time of evolution, we expect $S(\ell)/\ln 2$, with $\ell$ being the subsystem size,  to follow volume-law behavior, growing linearly with $\ell$ with a slope close to $1$, corresponding to an infinite-temperature state. 
In contrast, for an MBL system, $S(\ell)/\ln 2$ exhibits logarithmic growth in time, at $t\to\infty$ saturating at a value that is linear in $\ell$, but with a slope that is smaller than one and depends on the initial state~\cite{Serbyn2013, Bardarson2012}.

In our analysis, we compute the averaged entanglement entropy over all the subsystems of block size $\ell$ ($1\leq \ell \leq 8$) as a function of time $t$, up to the maximum simulation time $t_{\mathrm{f}}$.
As the systems we study are disordered, we would not like to restrict ourselves to the half-chain entanglement entropy, as this corresponds to the entanglement entropy of only the two subsystems with length $L/2$. Instead we are interested in studying the behavior of our disordered systems at a subsystem level, and extract entanglement entropy scaling results by analyzing subsystems of different sizes.
More specifically, 
for any fixed disorder realization we
compute the averaged entanglement entropy,
\begin{eqnarray}
\Sent(\ell) &=& -\tfrac{1}{N^{(\ell)}} \sum_{\kappa = x_i}^{x_f} \mathrm{Tr} (\rho^{(\ell)}_\kappa \ln \rho^{(\ell)}_\kappa)
\text{ .}
\end{eqnarray}
This averages over $N^{(\ell)}$ consecutive % different
block locations specified by the position $\kappa \in [x_i,x_f]$
of the first (left-most) spin within the block.
 To reduce finite size effects, we do not include % discard
$L/6$ spins on the very left and right of the chain
in the above average, thus $x_i=L/6+1$ and $x_f=5L/6-\ell+1$.

In \Fig{fig:caee}, we plot $\Sent(\ell)$ at the final time $t_{\mathrm{f}}$ as a function of $\ell$ for each disorder realization in different parameter regimes.
With $L=48$ at strong disorder [$\alpha=0.3$ in \Fig{fig:caee}(a),
or $\alpha=0.5$ in \Fig{fig:caee}(b)], we find that the majority of realizations lie between the volume law and area law, even after a long time, $t_{\mathrm{f}}=500$. This indicates that the system has not thermalized at these evolution times, signaling subthermal behavior.
On the other hand, at weaker effective disorder ($\alpha=1$) we observe that $\Sent$ approaches the volume law, with a slope close to $1$, for almost all realizations, already at a shorter time, $t=150$ [\Fig{fig:caee}(c)].

To further distinguish the subthermal and thermal behavior, we compute the distribution $P(\Sent)$ and its variance
$\sigma_{S}$ for $\ell=5$ at $t=t_{\rm f}$ [\Fig{fig:caee}(d,e)], for system size $L=48$ and different disorder strengths.
We note that the variance of the entanglement entropy of system's eigenstates has also been used previously to study disordered Ising models \cite{Kjall2014},
with maximum variance serving as an indicator of the MBL--thermal crossover in finite-size systems. In our case, we will focus on the entanglement entropy after the quench and the subthermal--thermal crossover.
We observe in \Fig{fig:caee}(d)
that at weaker disorder the distribution narrows and
$\sigma_{S}$ approaches zero with increasing $\alpha$, while for stronger disorder the distribution remains broad, with sizable variance $\sigma_{S}\gtrsim 0.1$ [\Fig{fig:caee}(e)], indicating subthermal behavior at studied system sizes.

To investigate the dependence of the (sub)thermal behavior on the system size, we perform entanglement entropy calculations for $L=24$ and $96$ and disorder strength $\alpha=0.5$.
In Figs. \ref{fig:caee}(f-h), clear subthermal behavior is evident at $t_{\mathrm{f}}=500$ for all system sizes.
However, increasing the system size enhances the tendency towards thermalization, reflected in the reduced variance of entanglement entropy, $\sigma_{S}$ [Fig.~\ref{fig:caee}(j)].

\begin{figure}[bt!]
\includegraphics[width=1\linewidth]{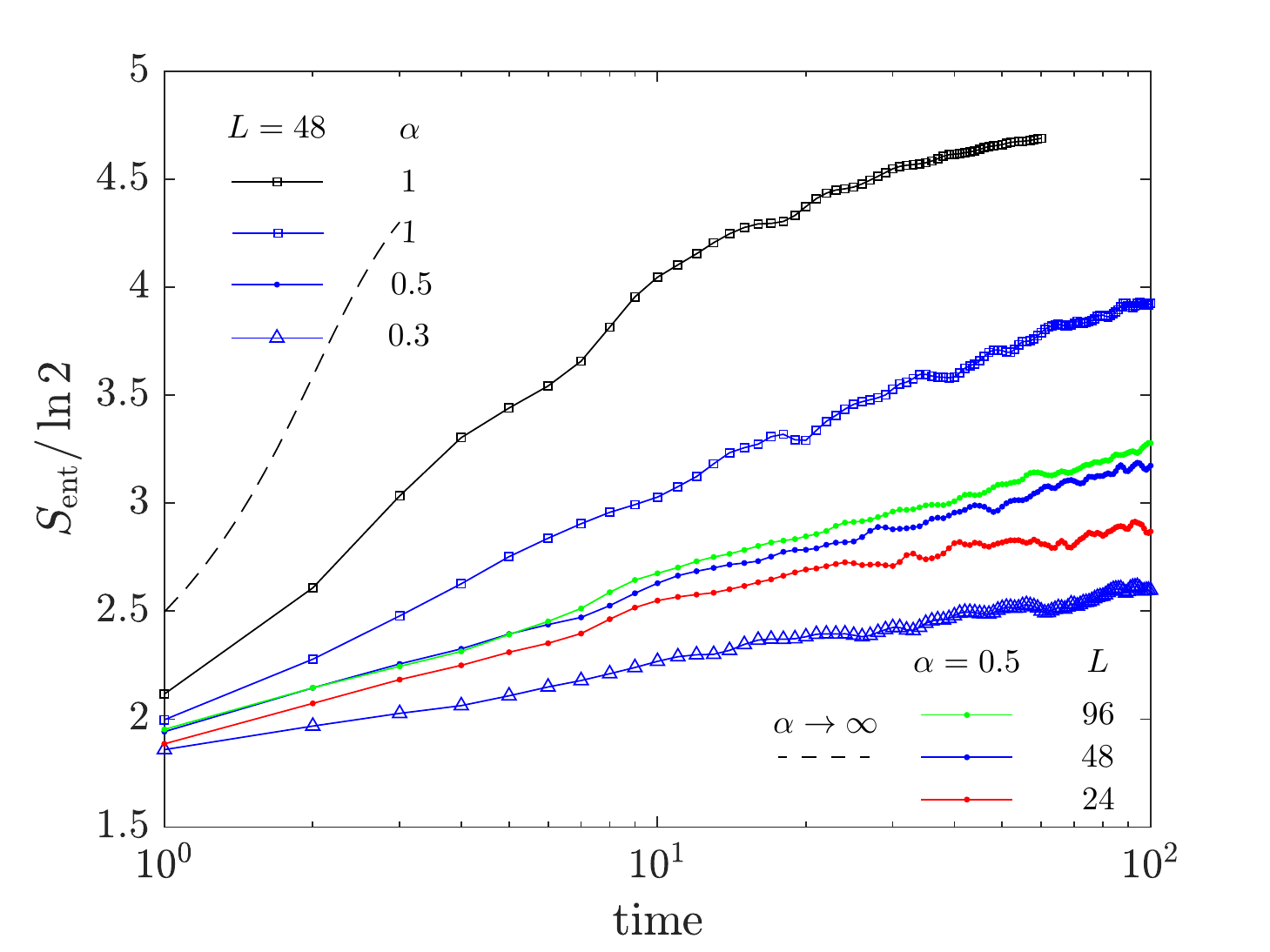}
\caption{Entanglement entropy $\Sent (\ell = 5) / \ln 2$ growth with time $t$ for the same parameter combinations as in Fig.~\ref{fig:caee},
averaged over the 25 (out of 100) realizations
with the lowest entanglement at $t_{\rm f}$. The black line refers to a single realization with $L=48$ and $\alpha=1$ that shows behavior very close to the volume law with a maximum possible prefactor,
$S_{\rm ent}^{(\ell)} / \ln 2
\leq \ell = 5$ (upper axis). For comparison, the dashed line corresponds to the entanglement growth in the uniform Heisenberg model ($\alpha\rightarrow\infty, J_i=1, \forall i$).
}
  \label{fig:entropy_growth}
\end{figure}

To analyze the entanglement entropy growth in time, we studied the time evolution of $\Sent (t)$ at $\ell=5$ for the parameter combinations discussed above (see Fig.~\ref{fig:entropy_growth}). The black data points show $\Sent (t)$ for a particular disorder realization for $\alpha=1$, that appears to obey the volume law, as seen in Fig.~\ref{fig:caee}(c).
Interestingly, the entanglement entropy grows logarithmically in time, in contrast to the linear behavior in conventional thermalizing systems.
Thus, for weaker disorder ($\alpha=1$), we expect that the system can reach a thermal state, although the dynamics are very slow.

To gain further insight into the structure of subthermal states, we investigate the entanglement growth of those disorder realizations, which do not reach the volume-law scaling at $t_{\rm f}$, and therefore show the most pronounced non-ergodic behavior.
To that end, we average $S_{\rm{ent}}(t)$ over the 25\% % $25$ (out of $100$)
disorder realizations with the lowest $S^{(5)}_{\rm{ent}}$
scaling at $t=500$ for the parameter combinations
in \Fig{fig:caee}, and the results are shown in \Fig{fig:entropy_growth}. We observe a logarithmic growth, $S_\mathrm{ent}\sim\log t$, of the entanglement entropy with time, for all the parameter cases corresponding to strong disorder.
At $L=48$ (blue lines), the entanglement entropy growth depends on the disorder strength, with the strongest disorder showing the slowest growth.
Additionally, at $\alpha=0.5$, we notice that the entanglement entropy of these realizations grows slightly faster when the $L$ is increased. Nevertheless, this drift with respect to system size is very slow, indicating the robustness of the subthermal regime at experimentally relevant times. Eventually, we expect the thermalization to take place; however, the dynamics are extremely slow, which does not allow us to precisely determine the very long thermalization time scale and required system sizes. 

\begin{figure}[bt!]
\includegraphics[width=0.8\linewidth]{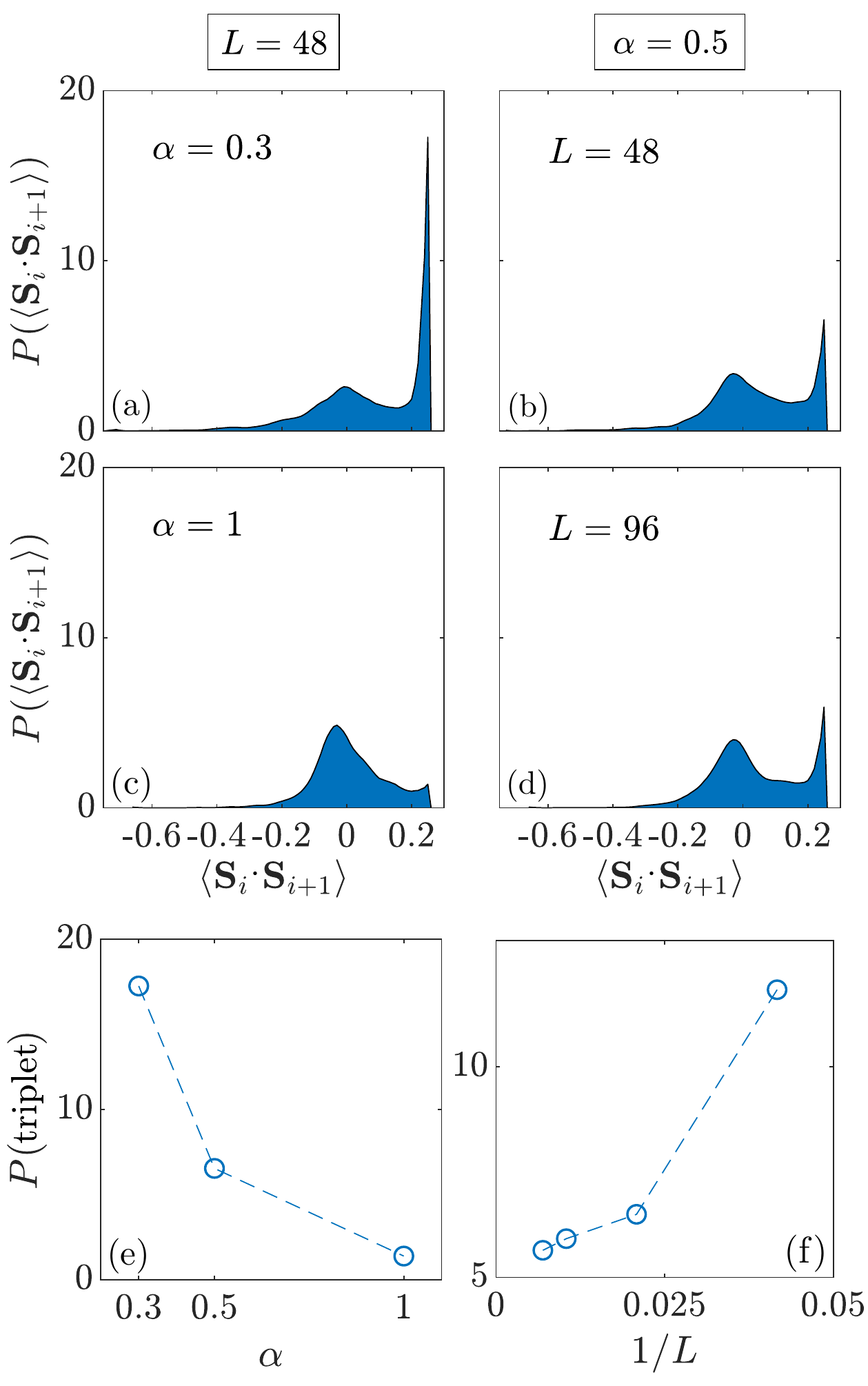}
 \caption{Distribution of nearest-neighbour spin-spin correlations at final time $t_{\mathrm{f}}$ for the odd bonds initialized as spin triplets of $100$ realizations for each combination of system size $L$ and disorder strength $\alpha$ shown.
 The data, collected vs.
 $x\equiv \langle\textbf{S}_i\textbf{S}_{i+1}\rangle$
 with uniform bin-size $dx = 0.01$, is shown here
 as normalized distributions $\int P(x) dx = 1$.
 By construction, at time $t=0$, $P(x)=\delta(x-\frac{1}{4})$.
 We chose $t_\mathrm{f}=500$ for all cases considered, except $\alpha = 1$, $L=48$, for which we used $t_\mathrm{f} =150$. (a-c): fixed $L=48$, $\alpha\in \{0.3, 0.5, 1  \}$. (d): $L=96$, $\alpha=0.5$ 
 (e, f): Percentage of bonds that remain localized in triplet states
 at final time $t_{\mathrm{f}}$ as a function of $\alpha$ or $1/L$,
 which due to the pinning size is equivalent here to $P(x)$ with $x\simeq
 (\mathbf{S}_{i}\cdot \mathbf{S}_{i+1})_{\rm triplet} = 0.25$,
 i.e., the average weight 
 in the last bin ($0.24 < x \leq 0.25$).
}
  \label{fig:corr_hist}
\end{figure}

\subsection{Spin-Spin correlations}

A defining feature of thermalization is that, in the course of its time evolution, the system loses the memory of its initial conditions.
That is, if a system is thermal, despite being in a pure state, after reaching a stationary state, values of local observables coincide with the thermal expectation values described by an appropriate Gibbs ensemble.
If the energy density of the initial state is set to be in the middle of the spectrum, $\braket{H}\sim 0$, the local reduced thermal density matrix will correspond to an infinite temperature limit $T \rightarrow \infty$.
Therefore, we expect the nearest-neighbor correlations to evolve towards $\braket{\mathbf{S}_{i}\cdot \mathbf{S}_{i+1}}\sim 0$ % =
in the long time limit.
On the other hand, in the MBL phase, certain local observables at long times will correlate with their initial values, reflecting the emergence of local integrals of motion. In SU$(2)$ subthermal disorder spin chains, we expect the spin-spin correlations for strongly coupled pairs of spins to be approximately conserved over time~\cite{Abanin2020}. Thus, we expect that some of the spin pairs initialized as triplets (half of the total pairs in the chain), which are coupled by \textit{strong} bonds, will retain the initial triplet correlations, $\braket{\mathbf{S}_{i}\cdot \mathbf{S}_{i+1}} \simeq 0.25$.

Figure \ref{fig:corr_hist} shows the distribution of the nearest-neighbor spin correlations obtained after a sufficiently long time ($t_f=500$),
focusing only on the spin pairs coupled by odd bonds that formed triplets at $t=0$. 
We consider systems of size $L=48$ with disorder strength $\alpha\in\{0.3, 0.5, 1\}$. For the disorder value $\alpha = 0.5$, we also study different system sizes, $L=\{24, 48, 96, 144\}$.
% It is clear that // AW
At $t_f=500$, we see 
strong evidence of ergodicity breaking at strong disorder ($\alpha = 0.3$), as the percentage of bonds remaining in the triplet state is significant [\Fig{fig:corr_hist}(e)], in agreement with the entanglement entropy results, discussed above. This tendency is reduced as the disorder becomes weaker. Similarly, at constant disorder strength $\alpha=0.5$,  % evidence
as the system size increases, the percentage of the initial triplet bonds that remain localized at $\langle\textbf{S}_i\textbf{S}_{i+1}\rangle
\simeq 0.25$ slowly decreases [\Fig{fig:corr_hist}(f)]. 
In particular, extrapolating this percentage to the infinite-size limit $1/L \to 0$, the percentage of localized bonds
 appears to remain above $P(0.25) \gtrsim 5\%$.
When compared to the bulk distributions in the upper panels
of \Fig{fig:corr_hist}, this would suggest that a peak
towards the right boundary at $x=\langle\textbf{S}_i\textbf{S}_{i+1}\rangle \simeq 0.25$ remains present and pronounced even for $L\to\infty$.
{This may be considered a consequence of so-called weak links
in the system [cf. discussion accompanying \Eq{eq:distribution}]
that lead to very slow convergence with system size in terms
of local (nearest-neighbor) expectation values.} 
The persistence of triplet correlations indicates the existence of approximate conservation laws, which correspond to a total spin of neighboring spin pairs coupled by $J$, which is much larger than their coupling to other neighbors~\cite{Abanin2020}.
Yet one can also observe the onset of a down-turn for the two left-most
data points in \Fig{fig:corr_hist}(f) which thus, eventually,
may further weaken the overall presence of localized bonds.
Overall, for strong disorder ($\alpha = 0.5$) the ergodicity-breaking behavior is system-size dependent, with a tendency towards very slow thermalization as $L$ increases, while for weaker disorder ($\alpha=1$) the evidence of thermalization is present already for small systems [Fig.~\ref{fig:corr_hist}(c)].

\begin{figure}[bt!]
\includegraphics[width=1\linewidth]{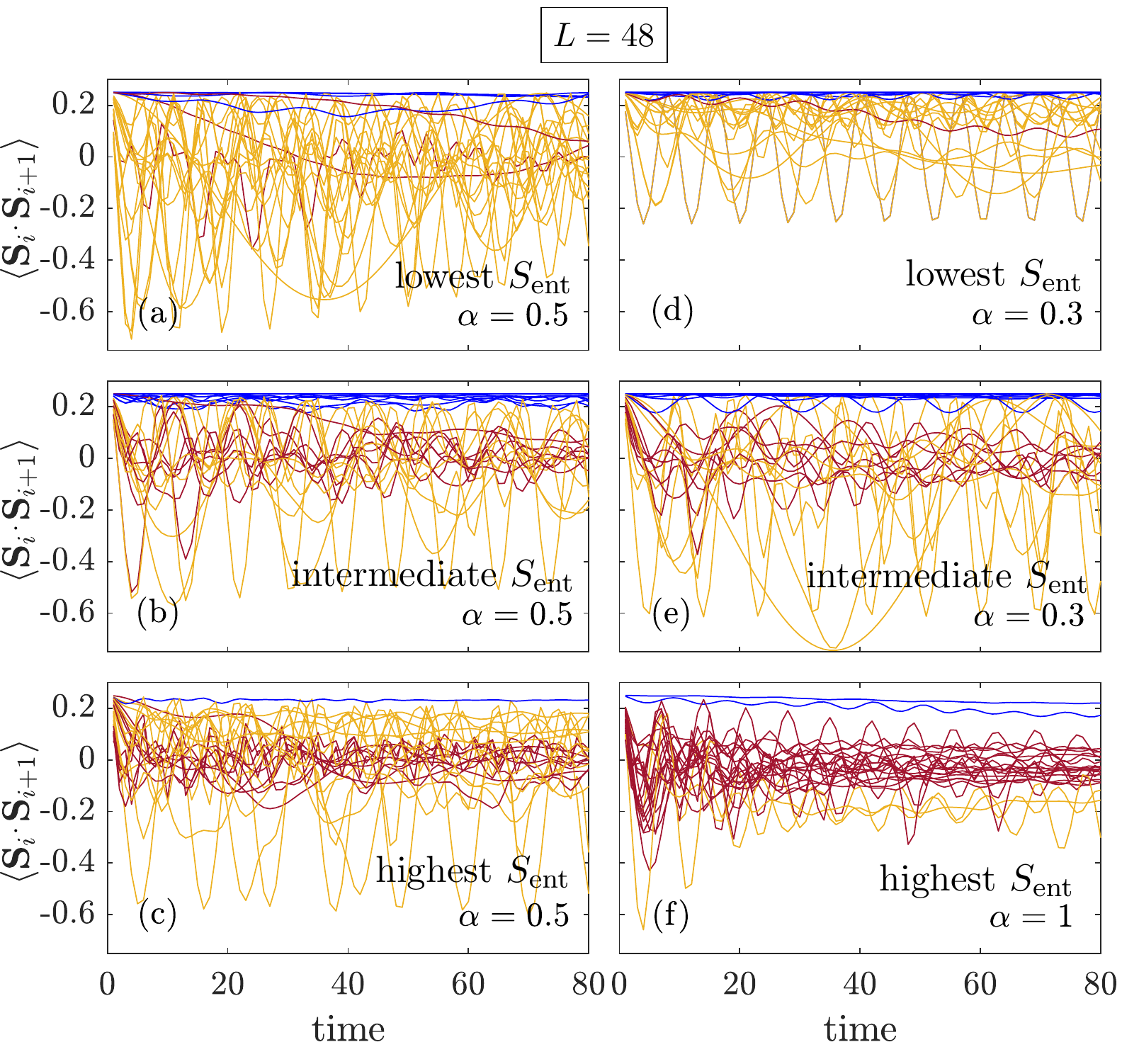}
\caption{Time evolution of $\langle \mathbf{S}_i \cdot \mathbf{S}_{i+1} \rangle$ of the odd bonds of single disorder
realizations of $L=48$ systems with different $\Sent$ behavior
as indicated top to bottom: the labels 
lowest, intermediate or highest $\Sent$ indicate that from all the disorder realizations in Fig.~\ref{fig:caee}(a-c)
for the specified $\alpha$ 
the ones were picked that have the lowest, intermediate, or highest $\Sent$ scaling, respectively, where % reordered sentence (no reason to label this as \new{..}
we define as lowest $\Sent$: $\Sent^{(8)}<4$, intermediate $\Sent$: $4 \leq \Sent^{(8)}\leq 6$ and highest $\Sent$: $\Sent^{(8)}>6$.
Blue (red, or yellow) lines correspond to
bonds that approach values of $\langle \mathbf{S}_i \cdot \mathbf{S}_{i+1}
\rangle$ close to the maximum $+0.25$ (around zero, or else), respectively.
}
\label{fig:corr_growth}
\end{figure}

So far, both the behaviour of the entanglement entropy and the spin-spin correlations at long times have shown evidence of a subthermal regime that is both disorder- and size-dependent. In order to understand the dynamics of this regime, Fig.~\ref{fig:corr_growth} illustrates the time evolution of the spin-spin correlations of the bonds initialized as spin triplets. Starting with the case $\alpha = 0.5$
and $L=48$ (left panels), we plot the time-evolution of $\braket{\mathbf{S}_i\cdot \mathbf{S}_{i+1}}$ for all the odd bonds (hence $\langle \mathbf{S}_i\cdot \mathbf{S}_{i+1}\rangle(t=0)
=0.25$)
of realizations with different entropy scaling (Fig.~\ref{fig:caee}). In particular, 
in \Fig{fig:corr_growth} the top middle, and bottom panels
correspond to disorder realizations from Fig.~\ref{fig:caee} having the lowest, intermediate, or the highest entanglement scaling, respectively. We can distinguish three different dynamical behaviours. Firstly, there are bonds with correlations
well conserved over time (blue lines),
suggesting emergent local integrals of motion. For realizations that exhibit higher entanglement entropy, a majority of bonds show correlations oscillating in the vicinity of $0$ (red lines), indicative of corresponding local observables reaching equilibrium. In all cases, there are also bonds with other non-trivial dynamics exhibiting oscillations of different frequencies and amplitudes (yellow lines).  This reflects the complex dynamics of the system due to strong non-uniformity of the couplings.

The case $\alpha=0.3$ [right panels in \Fig{fig:corr_growth}] shows similar phenomenology. The realization with the lowest entanglement entropy, Fig.~\ref{fig:corr_growth}(d), contains both bonds with conserved correlations and the ones showing non-trivial dynamics. For $\alpha=0.3$, we observe realizations with intermediate entropy scaling at most. Such a realization [Fig.~\ref{fig:corr_growth}(e)] shows similar dynamics as those in [Fig.~\ref{fig:corr_growth}(b)], as both of them have similar entropy scaling. On the other hand, for weaker disorder $\alpha=1$, for the realization with the highest entanglement entropy, the vast majority of bonds show oscillations around $0$ (thermal bonds), consistent with the entanglement entropy reaching a thermal volume law.

The results described above provide an intuitive picture of the dynamics in the subthermal regime, consistent with previous findings \cite{Potter2016, Protopopov2017, Abanin2020}. Specifically, the absence of thermalization within accessible evolution times stems from the existence of pairs of strongly coupled spins, the total spin of which is an approximate integral of motion. 

\section{SU$(3)$-symmetric model} 

Next, we consider the dynamics of a disordered SU(3)-symmetric
Heisenberg model. First, we need to 
understand the range of values expected for the spin-spin correlations. We assume
individual spins in the $N$-dimensional fundamental irreducible representation of SU$(N)$.
For an SU$(N)$ invariant state, the two-site density matrix of dimension $N^2\times N^2$ has a symmetric subspace
of dimension $d_S\equiv N(N+1)/2$ and an anti-symmetric subspace 
of dimension $d_A\equiv N(N-1)/2$.
Then the spin-spin correlations are given by
\begin{align}
 \langle \textbf{S}_i\cdot\textbf{S}_j \rangle = \tfrac{1}{2}\left(p_S - p_A -\tfrac{1}{N}  \right)
 \label{eq:spinspin2}
\end{align}
where $p_S$ and $p_A$ are the total weights of a two-spin state in the symmetric or antisymmetric subspace, respectively. 
For a thermal state at infinite temperature, 
$\langle \textbf{S}_i\cdot\textbf{S}_j \rangle
= \frac{1}{2}( \frac{d_S}{N^2} - \frac{d_A}{N^2}-\frac{1}{N})=0$
for any $N$.
For $N=3$, a fully symmetric state of two nearby spins has a correlation $\langle\textbf{S}_i\cdot\textbf{S}_j\rangle=+1/3$, while a fully anti-symmetric state has $\langle\textbf{S}_i\cdot\textbf{S}_j\rangle=-2/3$.

\begin{figure}[bt!]
\includegraphics[width=1\linewidth,trim = 0.0in 0.0in 0.0in 0.0in,clip=true]{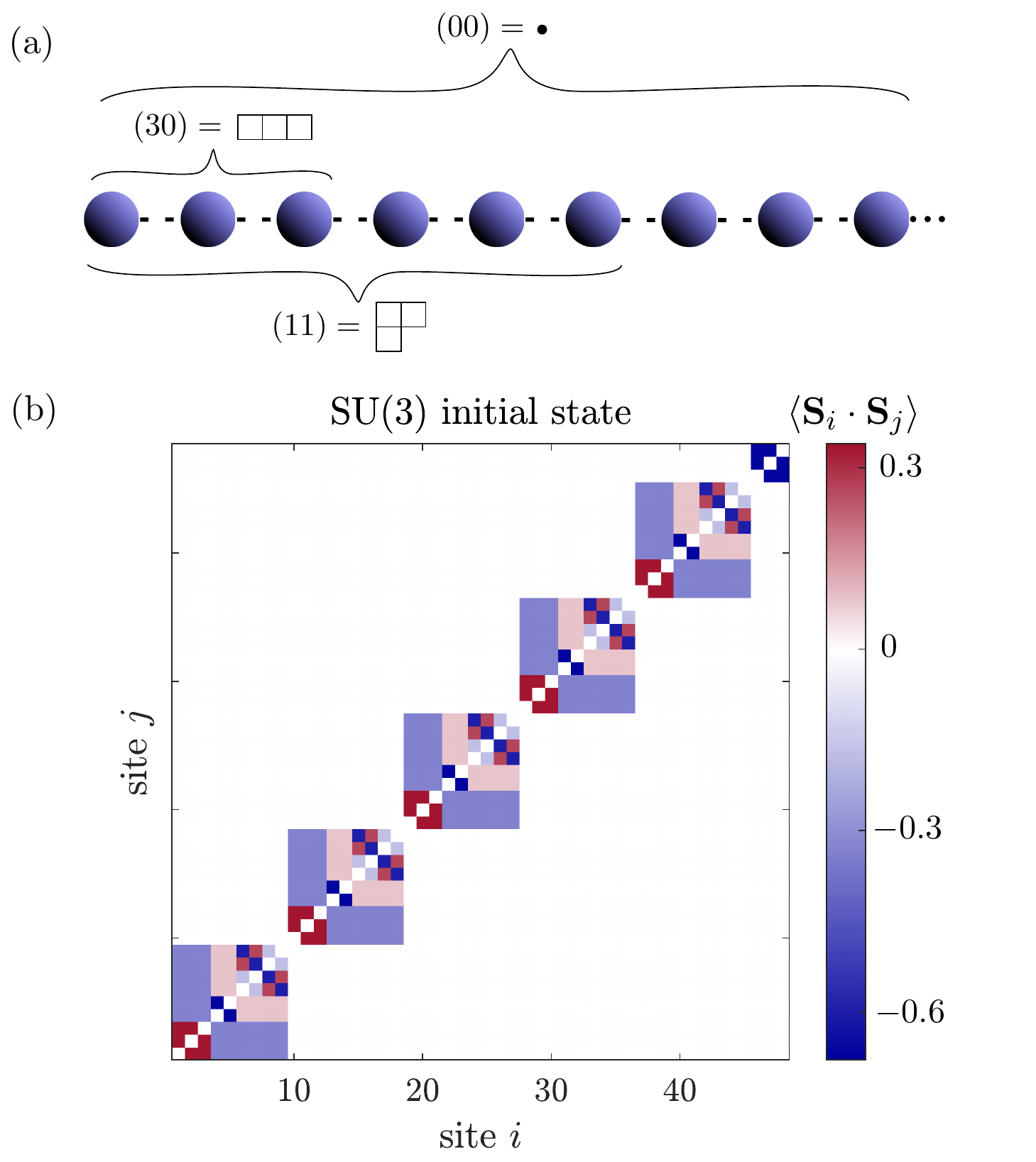}
  \caption{(a):  Schematic depiction of the structure of the initial state for the SU$(3)$ model.(b): Spin-spin correlations $\langle\textbf{S}_i\cdot\textbf{S}_j\rangle$ color-plots for the SU$(3)$ initial state. This state involves nearest neighbor bonds in both the fully symmetric and the fully-antisymmetric subspace having $\langle\textbf{S}_i\cdot\textbf{S}_j\rangle=+1/3$ (red) and $-2/3$ (blue), respectively.}
  \label{fig:initial_states_SU3}
\end{figure}

\subsection{SU$(3)$ model initial state}

Similar to the SU$(2)$ case, we choose an initial state with energy lying close to the middle of the spectrum ($\langle H \rangle\sim 0$).
The choice of the subspaces for each site is also restricted by the fact that the total MPS is again assumed to be in the singlet sector $(00)$.

In what follows we use the standard
multiplet labels for SU(3) based on Young tableaus
\cite{Young30,Cahn84}.
Specifically, SU(3) requires two labels for each multiplet,
$q\equiv(q_1,q_2)\equiv(q_1 q_2)$ which specify
the Young tableaux of two rows with $q_1+q_2$ and $q_2$
boxes in the first (second) row, respectively.
The defining representation
is thus ${\bf 3} \equiv (10)$, and its dual
$\bar{\bf 3} \equiv (01)$. The spin operators
transform in the adjoint representation ${\bf 8} = (11)$
which derives from $(10)\otimes (01) = (00) + (11)$,
with the $(00)\equiv {\bf 1}$ being the scalar singlet.
Furthermore, $(10) \otimes (10) = (20) + (01)$, with
$(20)\equiv  {\bf 6}$
the symmetric, and $(01) \equiv \bar{\bf 3}$
the antisymmetric subspace.

To construct the initial state, 
consecutive blocks of three sites are fused into the fully
symmetric multiplet (30), as shown in \Fig{fig:initial_states_SU3}(a). Three such neighboring blocks
are then fused into an overall singlet via an intermediate
pairing of two blocks into the (arbitrary but fixed)
adjoint (11) representation. As seen in
\Fig{fig:initial_states_SU3}(b), this state is not
symmetrical around the center of the chain, but neither are the randomized
Heisenberg couplings.  The last three spins on the right boundary of the chain form a bond in the fully anti-symmetric subspace
by themselves, as we wanted to keep the same system
sizes $L$ as in the SU$(2)$ analysis such that
we can also use the same disorder realizations
for the Heisenberg couplings $J_i$.  

The average energy of this initial state is around $E/L \sim -0.1$, 
tested for different random disorder configurations
at $\alpha=0.3$ and $0.5$. In Fig.~\ref{fig:initial_states_SU3}(b), the spin-spin correlations of the SU$(3)$ initial state is shown, revealing its structure. The colored boxes with spin-spin correlations values  $x\equiv {\langle\textbf{S}_i\cdot\textbf{S}_j\rangle=+1/3}$ (red) and $\langle\textbf{S}_i\cdot\textbf{S}_j\rangle=-2/3$ (blue) refer to pairs of spins initially in a fully symmetric and fully anti-symmetric subspace, respectively, which we will refer to as ``extremal" bonds below. 

\subsection{Results}

Similarly to our analysis of the SU$(2)$ model, we always
start with the same initial state as defined above [Fig. \ref{fig:initial_states_SU3})] which typically resides
close to the middle of the spectrum. For the case of SU(3),
we focus on $L=48$ throughout. We then apply the tDMRG time evolution with $dt=0.1$ Trotter time step keeping up to
$D^\ast = 1024$ SU(3) multiplets % kept during truncation
(corresponding to about $D\lesssim 18000$ states).
For the sake of a more direct comparison,
we employed the same set of disorder realizations (i.e. random $J_i$ couplings) 
for each parameter combination $(L, \alpha)$ in our SU$(3)$
study as for the SU$(2)$ case.

\begin{figure}[!http]
\includegraphics[width=1\linewidth]{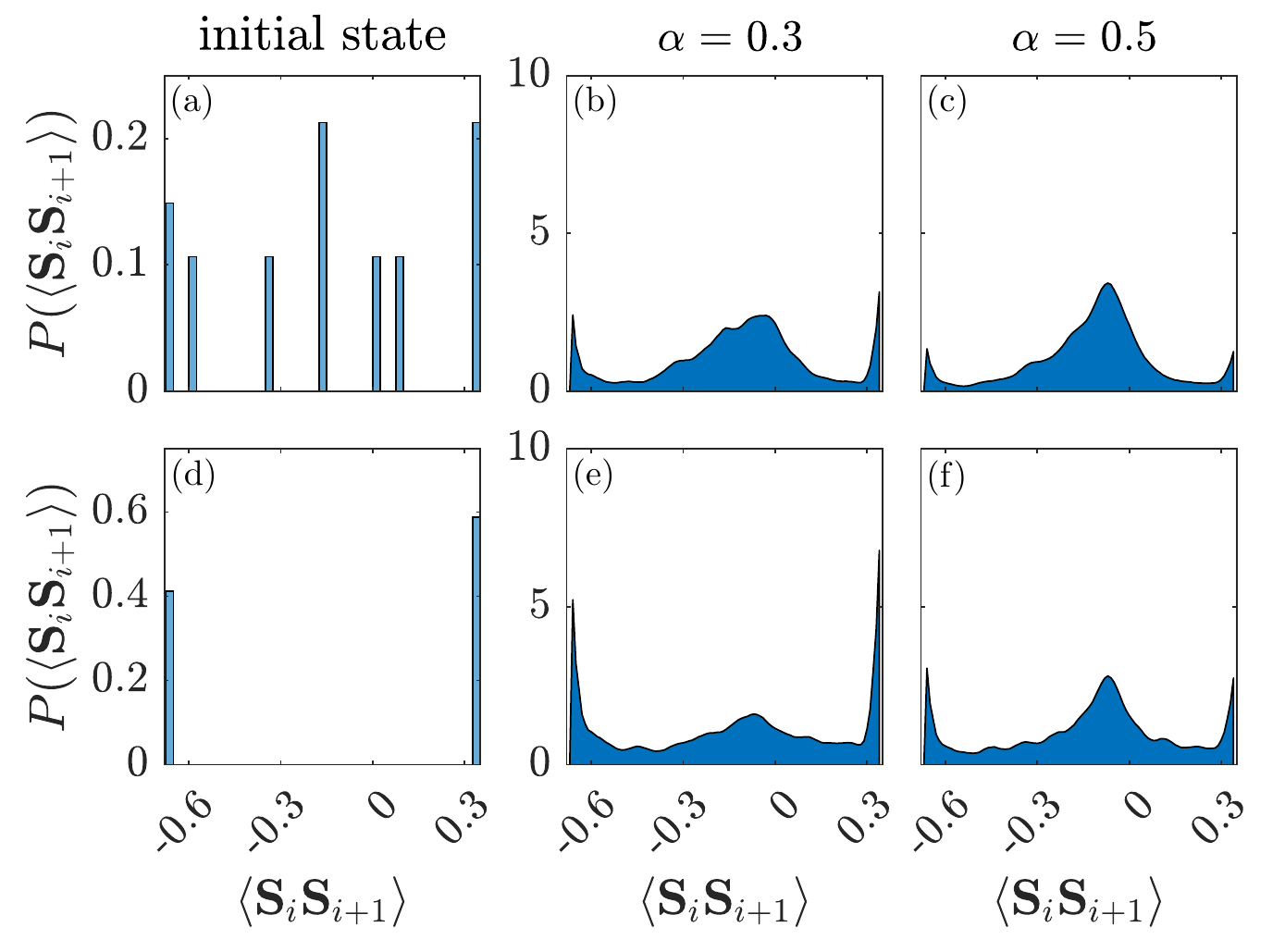}
\caption{Distribution of nearest-neighbour spin-spin correlations
for SU$(3)$ disordered systems of length $L=48$ 
after time evolution up to time $t_{\mathrm{f}} = 500$
with the initial state as in \Fig{fig:initial_states_SU3}
for all (or just the extremal initial) bonds 
in the upper panels (a-c) [lower panels (d-f)], respectively. 
Left panels (a,d): initial bond distribution at $t=0$.
Center panels (b,e): % (b):
$\alpha=0.3$.
Right panels (c,f): % (c):
$\alpha=0.5$.
The distribution $P(x)$ was obtained similarly to \Fig{fig:corr_hist},
with the same binning width $dx=0.01$. It is shown as a normalized
bar histogram in the left panels (with bars of width $dx$),
and as a normalized distribution function
in the center and right panels.
}
  \label{fig:histogramsSU3}
\end{figure}

\begin{figure}[!http]
\includegraphics[width=1\linewidth]{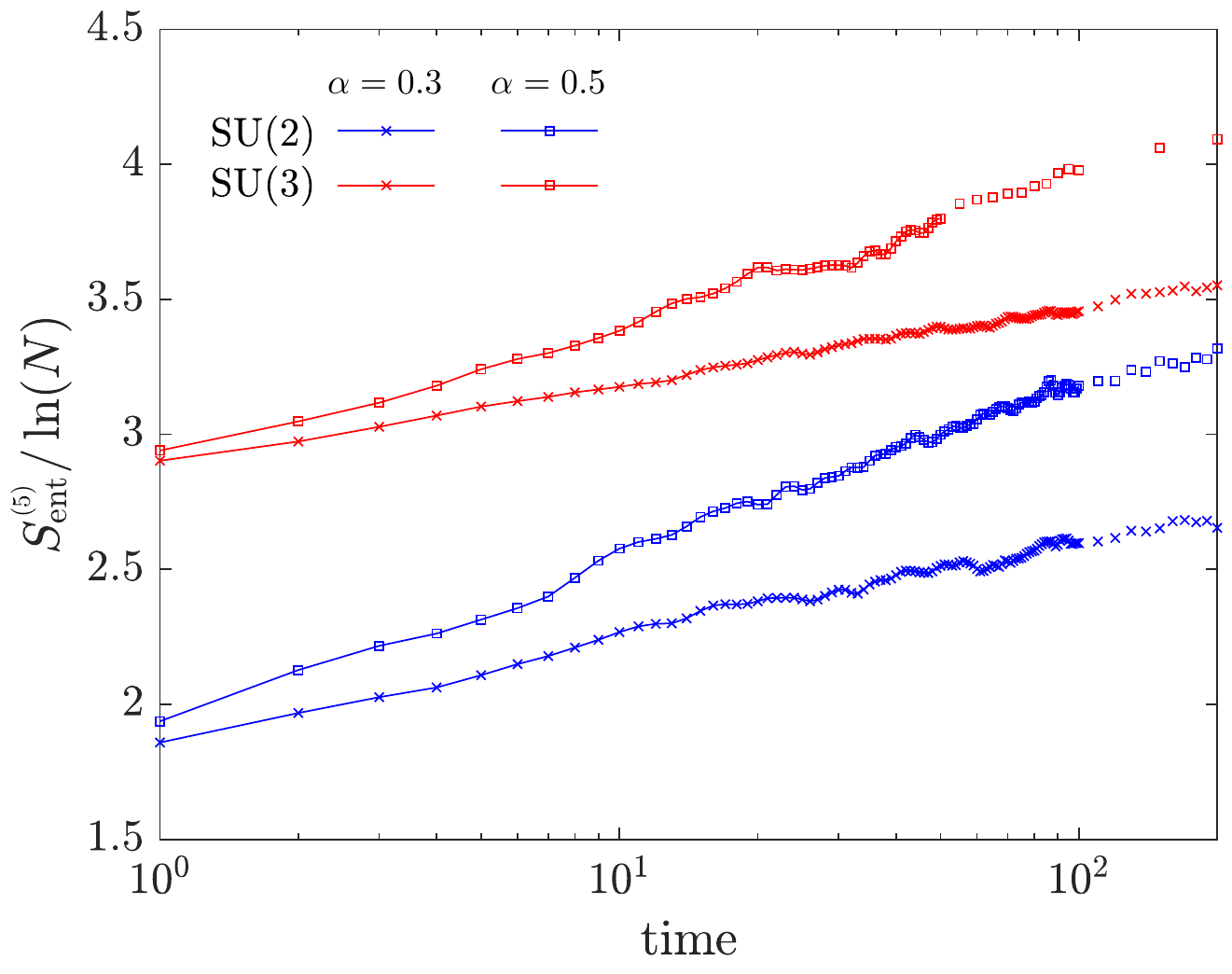}
\caption{Growth of the entanglement entropy of $\ell=5$ subsystems
with time, for SU$(2)$ and SU$(3)$ models and $L=48$, $\alpha=\{0.3, 0.5\}$.
The data is normalized by $\ln N$, thus having the common
upper bound $S_{\rm ent}^{(\ell)} / \ln N \leq \ell = 5$.
For each parameter combination,  % consider 
we focused on the 25\% 
realizations with the lowest entanglement entropy scaling
in Fig.~\ref{fig:caee}
as discussed with \Fig{fig:entropy_growth}.
}
  \label{fig:entropy_growthSU3}
\end{figure}

For the cases $\alpha=0.3, L=48$ and $\alpha=0.5, L=48$, for which the SU$(2)$ model showed non-ergodic behavior, we calculate the spin-spin correlations of the time-evolved states under the SU$(3)$-symmetric Hamiltonian, with the results shown in Fig.~\ref{fig:histogramsSU3}. % shows that
As seen from \Fig{fig:histogramsSU3}(a),
the percentage of bonds of the initial state that are localized in the fully-antisymmetric subspace ($x=-2/3$) is about 15\%, % $>10\%$
and in the fully-symmetric subspace ($x=1/3$) about $20\%$.
At $t_f=500$ this percentage is significantly reduced to
$<3\%$ 
[i.e., having $P(x) < 3$ given the binning width $dx=0.01$]
in \Figs{fig:histogramsSU3}(b,c)
for both cases of strong disorder.
Thus, the subthermal behavior, although still present,
is not as pronounced here, as in the SU$(2)$ case, suggesting that SU$(3)$ chains thermalize somewhat faster.

The upper panels in \Fig{fig:histogramsSU3} show
the bond distribution using the data from all bonds. It is also instructive to study the behavior of spin pairs initially in 
a fully symmetric (${\langle\textbf{S}_i\cdot\textbf{S}_j\rangle=+1/3}$) or antisymmetric (${\langle\textbf{S}_i\cdot\textbf{S}_j\rangle=-2/3}$) state [see \Fig{fig:initial_states_SU3}(a)].  
The distribution of spin-spin correlators at $t_{\rm f}$ for such ``extremal" bonds, which constitute approximately 35\% of all bonds,  is illustrated in the lower panels of \Fig{fig:histogramsSU3}. These plots demonstrate an enhanced probability for such spin pairs to retain their initial correlations at long times, indicating that dynamics are not fully ergodic at these evolution times.

Next, we study the dynamics of entanglement entropy. Figure \ref{fig:entropy_growthSU3} contrasts the growth of the entanglement entropy $\Sent (t) /\ln N$ 
in time for the SU$(3)$ case with with that for SU$(2)$ chains, for the same disorder configurations. Similar to \Fig{fig:entropy_growth},
 we chose the 
$25$\% realizations with 
the lowest $\Sent$ value reached at $t_{\rm f}$. The logarithmic growth of $\Sent$ following a quench for the SU$(3)$ model, evident in Fig.~\ref{fig:entropy_growthSU3}, is suggestive of a subthermal behavior.
Compared to the SU$(2)$ data
also included % results as
in \Fig{fig:entropy_growthSU3}, 
$\Sent /\ln N$
reaches a higher value for the 
SU$(3)$ case.  
This, however, can be attributed to the fact that the initial state we chose is more entangled for the SU$(3)$ case.
The slopes in the semilogarithmic plot in 
\Fig{fig:entropy_growthSU3}, on the other hand,
are comparable for $N=2$ with $N=3$ for the same value of $\alpha$. This entanglement entropy behavior indicates that disordered SU$(3)$-symmetric chains also exhibit non-ergodic dynamics up to relatively long time scales; at the same time, the above analysis of spin-spin correlations suggests that the SU$(3)$ chains show a somewhat stronger tendency to thermalization compared to the SU$(2)$ case.

\section{Conclusion}

In conclusion, we have studied the effect of SU$(N)$ symmetry on (absence of) thermalization in disordered spin chains. To that end, we investigated the real-time evolution of the SU$(N)$-symmetric disordered Heisenberg model in a quantum quench, starting from a symmetric weakly entangled state. Using matrix-product-states based methods exploiting SU$(N)$ symmetry, we were able to access long-time dynamics of large systems, with sizes well beyond those considered in previous studies using exact diagonalization. 

For the case of SU$(2)$ model, we found phenomenology consistent with the results of Ref.~\cite{Abanin2020}: at strong disorder, the system exhibits a subthermal regime, characterized by slow entanglement growth and absence of thermalization, attributed to the emergence of approximate integrals of motion, given by the total spin of pairs of strongly coupled neighboring spins. We investigated the distribution of $\Sent$ and correlation functions, finding them to be broad; the slow narrowing of these distributions with increasing system size suggests slow eventual thermalization. At weaker disorder, we observed thermalization evidenced by entanglement entropy reaching nearly maximum volume-law scaling. Interestingly, entanglement showed slow dynamics in this regime as well. Interestingly, in a recent paper \cite{mcroberts2023}, the classical equivalent of the disordered SU$(2)$-symmetric model was studied, where a regime of subdiffusive spin transport was observed.

Finally, we studied the SU$(3)$ symmetric model, not considered in previous papers, for the parameter regimes in which the SU$(2)$ model behaves in a non-ergodic way. The spin-spin correlations results indicated that the subthermal behavior, although present, is less pronounced, compared to the SU$(2)$ case. Similarly, the entanglement entropy growth remains logarithmic, but the $\Sent$ reaches values higher than the SU$(2)$ case for the same disorder strength and system size. Thus, we conclude that the phenomenology of SU$(3)$-symmetric chains is similar to the SU$(2)$ case, but with a stronger tendency to thermalization. 

Taken together with the resonance analysis within strong-disorder renormalization group for the SU(2) case~\cite{Abanin2020}, our results suggest that SU$(N)$-symmetric spin chains show eventual thermalization for arbitrary $N$ even at strong disorder. However, the subthermal regime is robust at $N=2,3$, showing a different entanglement pattern compared to either MBL or thermalizing phase. It would be interesting to observe the signatures of this nonergodic behavior in quantum simulation experiments. 

\section*{Acknowledgments}\mbox{}
Our numerical simulations employed the QSpace tensor library
\cite{Weichselbaum2012, Weichselbaum2020,qspace4u}
which can exploit general non-abelian symmetries in tensor network
algorithms. In this paper, this allowed us to fully exploit
the SU(2) and SU(3) system in our DMRG simulations.
This research
was funded in part (for JWL, JvD) by the Deutsche Forschungsgemeinschaft under Germany's Excellence Strategy EXC-2111
(Project No. 390814868), and is part of the Munich Quantum
Valley, supported by the Bavarian state government
with funds from the Hightech Agenda Bayern Plus.
A.W. was supported by the U.S. Department of Energy, Office of Science, Basic Energy Sciences, Materials Sciences and Engineering Division.
D. A. A. was supported by the European Research Council (ERC) under the European Union's Horizon 2020 research and innovation program (grant agreement No. 864597) and by the Swiss National Science Foundation. 

%%%%%%%%%%%%%%%%%%%%%%%%%%%%%%%%%%%%%%%%%%%%%%%%%%%%%%%%%%%%%%%%%%%%%%%%%%%%%%%%%%%%%

\appendix
\section*{Appendix}

In this appendix we provide information about the reliability of the simulations we performed for the analysis above. For that, we focus on the weaker disorder case with $\alpha=1$ and $L=48$. This is the case, which shows entanglement entropy scaling closer to the volume law, as shown in \Fig{fig:caee}(c), thus it would be wise to ensure that the time-evolution results are not affected by computational errors due to significant truncation.

The most instructive accuracy measure in DMRG is the reduced density matrix discarded weights \cite{Schollwock2005}:
\begin{align}
 \epsilon = 1 - \sum_{k=1}^M w_k
 \label{eq:discarded_weights}
\end{align}
\noindent where $w_k$ are the reduced density matrix eigenvalues and $M$ is the number of the dominant eigenvalues kept for the calculation. 

\begin{figure}[http!]
\includegraphics[width=1\linewidth]{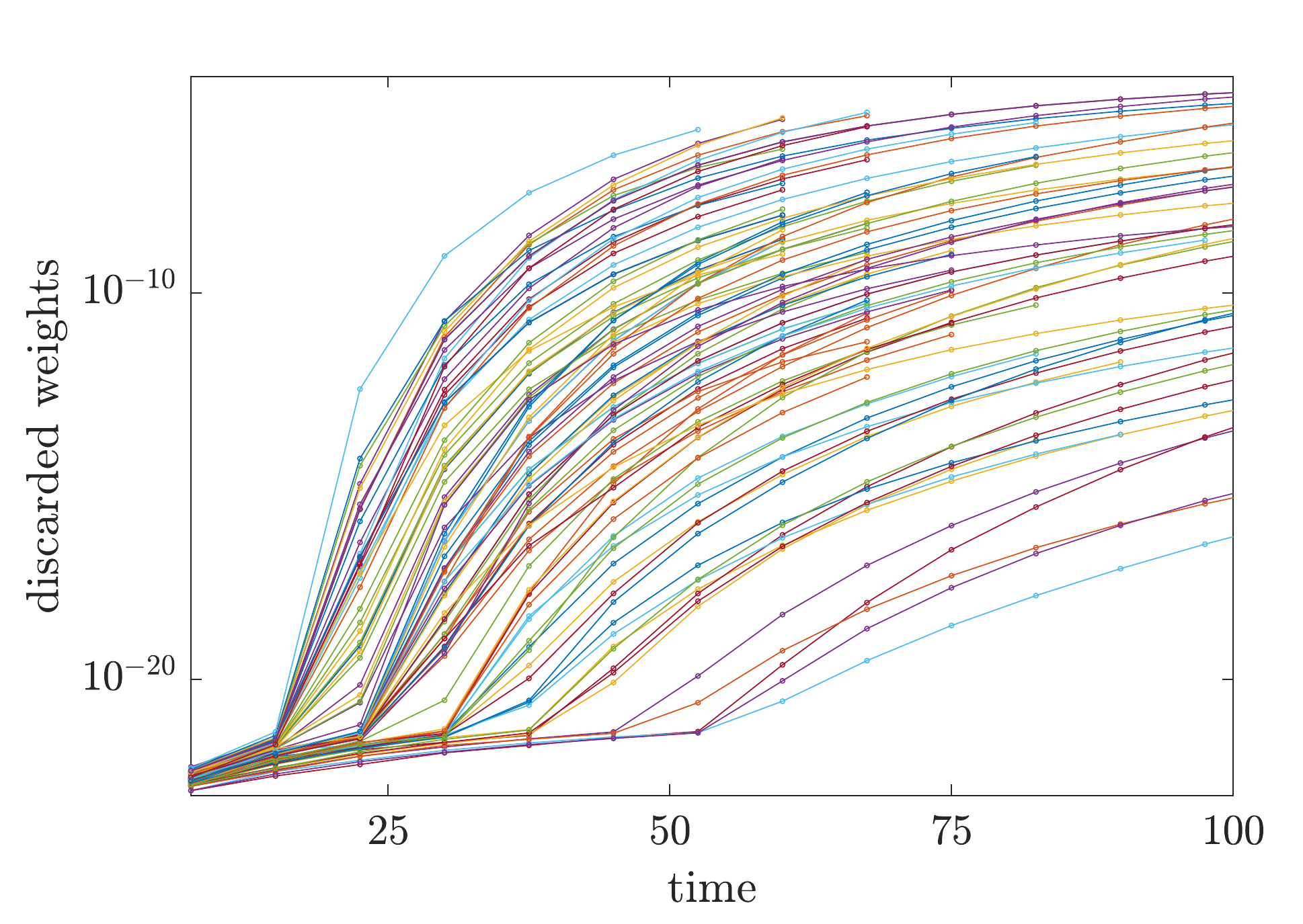}
\caption{Time evolution of the discarded weights for the case of $\alpha=1$ disorder strength and $L=48$.
}
  \label{fig:discarded_weights}
\end{figure}

In \Fig{fig:discarded_weights}, we show the time evolution of the discarded weights $\epsilon$ for the realizations with parameters $\alpha=1,~ L=48$. The discarded weights of some of the simulations have reached values of order $10^{-6} - 10^{-5}$, which is the reason that we show results only till $t_f=150$ in \Fig{fig:caee}(c) for this particular case. This is, nevertheless, in agreement with the point we make about the weaker disordered SU$(2)$-symmetric model: there is a tendency towards thermalization already at earlier times than the rest of the parameter cases, however with a logarithmic growth of the entanglement entropy, as seen in \Fig{fig:entropy_growth}.

\bigskip

\bibliography{SUn_references}

\end{document}